\documentclass[prl,aps,10pt,nofootinbib,twocolumn,superscriptaddress,preprintnumbers,balancelastpage,longbibliography]{revtex4-1}

\usepackage{dcolumn}
\usepackage[table]{xcolor}
\usepackage{youngtab}
\usepackage{ytableau}

\usepackage{booktabs}    % some options for tabulars\\
\usepackage{placeins}
\usepackage{soul}

\definecolor{redd}{rgb}{0.8, 0.1,0.2}
\definecolor{navy}{rgb}{0.05, 0.23,0.75}
\usepackage[colorlinks=true,citecolor=red,linkcolor=blue]{hyperref}
\usepackage{chngcntr}
\hypersetup{
     colorlinks   = true,
     citecolor    = navy,
	linkcolor = redd,
	urlcolor=navy,
	anchorcolor=blue
}
\usepackage{bbm}
\usepackage{amsfonts}
\usepackage{amsmath,amssymb}
\usepackage{mathrsfs}
\usepackage{epsfig}
\usepackage{graphicx}
\usepackage{wrapfig}                 
\usepackage{url}
\usepackage{hyperref}
\usepackage{float}
\usepackage{color}
\usepackage{multirow}
\usepackage{lipsum}
\usepackage{enumitem}
\usepackage{ctable}
\newcolumntype{L}{>{\centering\arraybackslash}m{1.5cm}}

\usepackage{enumitem}
\usepackage{chngcntr}

\usepackage{makecell}

\newcommand{\be}{\begin{equation}}
\newcommand{\ee}{\end{equation}}
\newcommand{\bea}{\begin{eqnarray}}
\newcommand{\eea}{\end{eqnarray}}
\newcommand{\bc}{\begin{center}}
\newcommand{\ec}{\end{center}}

\newcommand{\ZZ}{\mathbb{Z}}

\begin{document}
		
\title{
A General Prescription for Spurion Analysis of Non-Invertible Selection Rules
}

\author{Ling-Xiao Xu}
\email{phy.lingxiao.xu@gmail.com}
\affiliation{Abdus Salam International Centre for Theoretical Physics, Strada Costiera 11, 34151, Trieste, Italy}

\begin{abstract}
We formulate a general prescription for spurion analysis in particle-physics models whose selection rules are described by commutative non-invertible fusion algebras. The construction applies to fusion algebras containing non-invertible basis elements that need not be self-conjugate, thereby allowing us to systematically track coupling constants in arbitrary particle scattering processes at tree and loop orders, but without assuming faithful realization of the fusion algebra, or no other quantum numbers for dynamical particles. This unifies and streamlines the previous analysis of near-group fusion algebras and of the $\mathbb{Z}_M/\mathbb{Z}_2$ fusion algebras, and supports the broader viewpoint that the non-invertible selection rules often admit auxiliary descriptions using lifted Abelian groups with a structured set of explicit breaking terms.
\end{abstract}

\maketitle
%\tableofcontents

%%%%%%%%%%%%%%%%%%%%%%%%%%%%%%%%%%%%%%%%%%%%%%%%%%%%%%%%%
\section{Introduction}
\label{sec:intro}
%%%%%%%%%%%%%%%%%%%%%%%%%%%%%%%%%%%%%%%%%%%%%%%%%%%%%%%%% 
It has long been appreciated that \emph{Symmetry dictates interactions}~\cite{10.1093/nsr/nwz113}. This idea has become even more powerful with the recent generalizations of global symmetry in Quantum Field Theory (QFT), following the seminal work~\cite{Gaiotto:2014kfa}; see~\cite{Cordova:2022ruw, McGreevy:2022oyu, Gomes:2023ahz, Schafer-Nameki:2023jdn, Brennan:2023mmt, Luo:2023ive, Shao:2023gho, Costa:2024wks, Iqbal:2024pee, Kaidi:2026urc} for reviews.

Of particular interest in particle physics is the use of \emph{spurious symmetries} to understand small coupling constants and their stability under renormalization-group flow~\cite{tHooft:1979rat}. The basic idea is to promote coupling constants to nondynamical background fields carrying compensating charges. 
The purpose of this Letter is to extend this logic from ordinary group-based selection rules to selection rules governed by a commutative non-invertible fusion algebra. For brevity, we refer to them as non-invertible selection rules (NISRs).

We argue that NISRs are ubiquitous in particle physics. From a bottom-up perspective, the ordinary irreducible representations of a non-Abelian group already form a non-invertible fusion algebra under tensor product, without requiring the particles themselves to furnish those representations. From a top-down perspective, NISRs can also emerge naturally in the low-energy limit of string theory and may therefore serve as probes of the underlying ultraviolet structure~\cite{Kaidi:2024wio, Heckman:2024obe, Hamidi:1986vh, Font:1988nc, Kobayashi:1995py, Kobayashi:2011cw, Kobayashi:2025ocp, Dong:2026iwa}. These considerations motivate the need for a systematic spurion analysis of NISRs.

Our goal is practical. Given a fusion algebra organizing the allowed scattering channels of dynamical particles, we ask how to assign consistent labels to arbitrary interaction vertices~\cite{Suzuki:2025bxg, Suzuki:2025kxz}. The obstruction is immediate: non-invertible basis elements need not admit inverses, and fusion products may contain multiple channels.
As a result, the selection rule is no longer encoded by ordinary charge conservation in the group-theoretic sense. This makes it natural to trade the non-invertible structure for an auxiliary Abelian description in which the failure of ordinary charge conservation is captured by a structured set of explicit breaking terms.
We therefore adopt the viewpoint:
\\[0.25cm]
\textit{Many NISRs admit an auxiliary lifted description in terms of an Abelian group $G_{\mathrm{lift}}$ together with a structured set of explicit breaking terms}. 
\\[0.4cm]
This lifted group is not the symmetry of the system itself, but a bookkeeping device for organizing spurion labels. Different lifted descriptions may encode the same NISRs, and this nonuniqueness should be viewed as a property of the auxiliary description rather than as a physical ambiguity.

We formulate a general prescription directly from the fusion-algebra data. It applies to commutative fusion algebras with conjugation, including cases in which non-invertible basis elements are not self-conjugate. 
Going beyond the special classes previously studied in Refs.~\cite{Suzuki:2025bxg, Suzuki:2025kxz}, which relied on simplifying assumptions, the present framework turns the underlying insight into a general and systematic algorithm.
Our spurion analysis does not require the fusion algebra to be faithfully realized, nor does it assume that the fusion-algebra labels capture all quantum numbers of the particles.
In this sense, it complements the groupification perspective of~\cite{Kaidi:2024wio}, which, under its natural assumptions of faithful realization and the absence of additional quantum numbers, emphasizes the pattern of increasingly radiative breaking of NISRs at higher loop orders. 
By contrast, the main advantage of spurion analysis is that it furnishes a uniform consistency check for arbitrary processes, whether allowed or forbidden, thereby making the framework amenable to systematic phenomenological applications~\cite{Kobayashi:2024yqq, Funakoshi:2024uvy, Kobayashi:2024cvp, Kobayashi:2025znw, Suzuki:2025oov, Liang:2025dkm, Kobayashi:2025ldi, Kobayashi:2025cwx, Kobayashi:2025lar, Nomura:2025sod, Nomura:2025yoa, Chen:2025awz, Okada:2025kfm, Kobayashi:2025thd, Jangid:2025krp, Kobayashi:2025rpx, Jangid:2025thp, Nomura:2025tvz, Okada:2025adm, Nakai:2025thw, Okada:2026gxl, Kashav:2026jjg, Dong:2026crl, Nomura:2026hli, Okada:2026iob, Okada:2026bpp, Kobayashi:2025wty}.

%%%%%%%%%%%%%%%%%%%%%%%%%%%%%%%%%%%%%%%%%%%%%%%%%%%%%%%%
\section{General prescription}
\label{sec:generalities}
%%%%%%%%%%%%%%%%%%%%%%%%%%%%%%%%%%%%%%%%%%%%%%%%%%%%%%%%

Let $\mathcal{A}=\{1,x,y,\cdots\}$ denote the set of basis elements of a finite commutative fusion algebra, 
\bea
x\otimes y=\sum_z N^z_{x,y} z\ , 
\eea
where $N^z_{x,y}\in \mathbb{Z}_{\geq 0}$ are the structure constants. We assume that the algebra is equipped with an involutive conjugation $x\mapsto \bar{x}$ such that $1\prec x\otimes \bar x$ for every $x\in \mathcal{A}$, and such that conjugation is compatible with fusion. In the commutative case considered here, this means
\be
N^z_{x,y}=N^{\bar{z}}_{\bar{x},\bar{y}}, 
\ee
or equivalently, $\overline{x\otimes y}=\bar{x}\otimes \bar{y}$~\cite{Kaidi:2024wio}. Elements satisfying $x=\bar{x}$ are called self-conjugate.

The steps for determining the spurion label of a coupling are as follows.

\textit{(i) Fusion order.} Define the \emph{fusion order} $d_x$ of $x$ as the smallest positive integer such that 
\be
1\prec x^{d_x}. 
\ee
This is the analog of group-theoretical order, except that only the identity channel is required to appear. Since conjugation is compatible with fusion, one has $\overline{x^n}=\bar x^{\,n}$, and therefore $1\prec x^n \iff 1\prec \bar x^{\,n}$. It follows that 
\be
d_x=d_{\bar{x}}.
\ee

\textit{(ii) Cyclic reconstruction.}
The reconstruction problem is to determine whether the non-identity basis elements can be assigned residues in one or more ambient cyclic factors so that conjugation and fusion are represented consistently at the level of residue arithmetic. We therefore introduce candidate cyclic factors $\mathbb{Z}_{L_1},\mathbb{Z}_{L_2},\cdots,\mathbb{Z}_{L_m}$ and assign to each basis element $x$ a residue
\begin{equation}
r_\alpha(x)\in \mathbb Z_{L_\alpha},
\qquad \alpha=1,\dots,m,
\label{eq:residue}
\end{equation}
with $r_\alpha(x)\neq 0$ if $x$ is accommodated by the factor $\mathbb Z_{L_\alpha}$ and $r_\alpha(x)=0$ otherwise. For a given factor $\mathbb Z_L$, let $\Omega_L\subset \mathbb Z_L\setminus\{0\}$ denote the occupied residues. We restrict to injective assignments on $\Omega_L$, so that each occupied residue carries a unique basis element. Consistency then requires:
\begin{enumerate}
\item Order compatibility, i.e., 
\begin{equation}
\frac{L}{\gcd[r(x),L]}=d_x.
\label{eq:order-comp}
\end{equation}
Hence, a candidate factor $\mathbb{Z}_L$ can accommodate only basis elements whose fusion orders divide $L$.

\item Conjugation compatibility. At the pairwise level, $r(x)+r(y)\equiv 0\pmod L$ requires $1\prec x\otimes y$.
Because the identity channel singles out conjugate pairs, i.e., $1\prec x\otimes y \iff y=\bar x$, it implies 
\begin{equation}
r(\bar x)\equiv -\,r(x)\pmod L.
\label{eq:conj-comp}
\end{equation}
In particular, if $x=\bar x$, then $2\,r(x)\equiv 0\pmod L$, so a self-conjugate element can occupy only an order-two residue.

\item Fusion compatibility. At the pairwise level,
\begin{equation}
r(x)+r(y)\equiv r(z)\pmod L
\label{eq:pairwise1}
\end{equation}
with occupied $r(z)$ requires
\begin{equation}
z\prec x\otimes y.
\label{eq:pairwise2}
\end{equation}
These pairwise conditions are generally not sufficient. A cyclic reconstruction is accepted only if the candidate assignment satisfies the full modular consistency condition stated in the Supplementary Material. Because the fusion algebra need not arise from a group, the occupied subset $\Omega_L$ need not contain all nonzero residues of $\mathbb Z_L$.

\end{enumerate}

The non-group-like character of the non-invertible fusion algebra is encoded in fusion terms that connect different circles. Accordingly, the modular consistency condition is imposed only circle by circle, rather than on fusion products involving elements from different circles. In the lifted description, $\ZZ_L$ itself is explicitly broken when a fusion product $x\otimes y$, as in Eq.~\eqref{eq:pairwise2}, contains multiple basis elements accommodated in the same $\ZZ_L$.

\textit{(iii) Lifted Abelian group.}  
Each consistent cyclic reconstruction defines one cyclic factor, and their product gives the lifted Abelian group
\begin{equation}
G_{\mathrm{lift}}=\prod_{\alpha=1}^{m}\mathbb Z_{L_\alpha}.
\label{eq:glift}
\end{equation}
The lifted charge of a basis element is then the residue vector
\begin{equation}
q(x)=\bigl(r_1(x),\dots,r_m(x)\bigr)\in G_{\mathrm{lift}}.
\label{eq:charge}
\end{equation}
The factors of \(G_{\mathrm{lift}}\) are reconstructed circle by circle, whereas fusion terms connecting different circles appear as explicit breaking terms in the lifted description.

\textit{(iv) Spurion label.}
A vertex
\begin{equation}
V=\prod_{x\in \mathcal{A}} x^{m_x},
\qquad
m_x\in \mathbb Z_{\ge 0},
\label{eq:vertex}
\end{equation}
is allowed by the NISRs if
\begin{equation}
1\prec \prod_{x\in \mathcal{A}} x^{m_x}.
\label{eq:allowed-vertex}
\end{equation}
For such an allowed vertex, remove conjugate pairs $x\bar x$ to obtain the reduced vertex $R(V)$.
Because conjugate pairs carry opposite lifted charges, their total contribution vanishes. The associated spurion $\lambda_V$ is therefore assigned the compensating charge
\begin{equation}
q(\lambda_V)=-\,q\!\left(R(V)\right).
\label{eq:spurion}
\end{equation}
Some vertices allowed by the NISRs, namely those satisfying Eq.~\eqref{eq:allowed-vertex}, may nevertheless carry nonzero lifted charge~\cite{Suzuki:2025bxg, Suzuki:2025bxg, Suzuki:2025oov}. In this sense, $G_{\mathrm{lift}}$ is not a symmetry of the physical system, but is explicitly broken by such couplings. Conversely, not every $G_{\mathrm{lift}}$-breaking term is compatible with Eq.~\eqref{eq:allowed-vertex}. Therefore, NISRs correspond to a structured subset of $G_{\mathrm{lift}}$-breaking terms.

Amplitudes built from the elementary vertices then inherit their spurion charges automatically. Lifted charges are added under the gluing of vertices, while each internal contraction pairs a basis element with its conjugate and hence contributes zero net lifted charge. As a result, both tree-level and loop-level amplitudes inherit their spurion charges directly from the elementary couplings, with no further assignment required. The prescription can therefore be applied directly to any allowed coupling of interest, without first enumerating all allowed couplings.

%%%%%%%%%%%%%%%%%%%%%%%%%%%%%%%%%%%%%%%%%%%%%%%%%%%%%%%%%%%%%%%%%%%%%%%%
\section{Case studies}
\label{sec:examples}
%%%%%%%%%%%%%%%%%%%%%%%%%%%%%%%%%%%%%%%%%%%%%%%%%%%%%%%%%%%%%%%%%%%%%%%%

Spurion analysis for NISRs was previously developed in~\cite{Suzuki:2025bxg, Suzuki:2025kxz} under simplifying assumptions:
\begin{itemize}
\item Ref.~\cite{Suzuki:2025bxg} considered near-group fusion algebras $G+n^\prime$ with $n^\prime\in \ZZ_{\geq 0}$, built from an Abelian group $G$ together with a single self-conjugate non-invertible element $\rho$~\cite{Evans:2012ta}. The resulting lifted description was found to be $G\times \ZZ_2$, with explicit breaking terms arising from the self-couplings of $\rho$ when $n^\prime>0$, and the couplings between $\rho$ and the group elements of $G$.
\item Ref.~\cite{Suzuki:2025kxz} studied the $\mathbb{Z}_M/\mathbb{Z}_2$ fusion algebras, whose fusion rules take the form $[g^i]\otimes [g^j]=[g^{i+j}]+[g^{i-j}]$. These algebras may be viewed as a discrete avatar of the so-called cosine symmetry~\cite{Shao:2023gho}, since $[g^j]\sim 2\cos\left(\frac{2\pi}{M}j\right)$ reproduces the same multiplication law. All nontrivial elements are non-invertible except $j=0$ and, when $M$ is even, $j=M/2$, and all elements are self-conjugate because $1\prec [g^j]^2$. The lifted group is then a product of $\ZZ_2$ factors, one for each non-identity element, with the $\mathbb{Z}_M/\mathbb{Z}_2$ fusion rules encoded by explicit breaking terms.
\end{itemize}
In the present framework, these results are recovered straightforwardly from the general prescription. 

By contrast, for a general non-invertible fusion algebra, we do not assume an underlying group structure or self-conjugate non-invertible elements. Applying the general prescription, we analyze a broad class of examples up to rank 8 from~\cite{Kaidi:2024wio, Dong:2025jra}; the results are summarized in the Supplementary Material, in particular in Tables~\ref{tab:KTZ} and~\ref{tab:Dong}. 
We now illustrate the general prescription in detail using the fusion algebra shown in Table~\ref{tab:fusion-C9}. The remaining examples can be treated analogously.

\begin{table}[t]
\centering
\caption{Fusion rules of the algebra (C.9) from~\cite{Kaidi:2024wio}.}
\label{tab:fusion-C9}
\setlength{\tabcolsep}{6pt}
\renewcommand{\arraystretch}{1.15}
\begin{tabular}{c|cccccc}
 $\otimes$ & $1$ & $2$ & $3$ & $4$ & $5$ & $6$ \\
\hline
$1$ & $1$ & $2$ & $3$ & $4$ & $5$ & $6$ \\
$2$ & $2$ & $3$ & $1$ & $5$ & $6$ & $4$ \\
$3$ & $3$ & $1$ & $2$ & $6$ & $4$ & $5$ \\
$4$ & $4$ & $5$ & $6$ & $ 3\!+\!4\!+\!6$ & $ 1\!+\!4\!+\!5$ & $ 2\!+\!5\!+\!6$ \\
$5$ & $5$ & $6$ & $4$ & $ 1\!+\!4\!+\!5$ & $ 2\!+\!5\!+\!6$ & $ 3\!+\!4\!+\!6$ \\
$6$ & $6$ & $4$ & $5$ & $ 2\!+\!5\!+\!6$ & $ 3\!+\!4\!+\!6$ & $ 1\!+\!4\!+\!5$ \\
%\bottomrule
\end{tabular}
\end{table}

As an illustrative example, let us consider the fusion algebra in Table~\ref{tab:fusion-C9}. The basis elements $1, 2, 3$ are invertible, whereas $4, 5, 6$ are non-invertible. The conjugation action is
\be
\bar{1}=1, \quad \bar{2}=3, \quad \bar{3}=2, \quad \bar{4}=5, \quad \bar{5}=4, \quad \bar{6}=6\;.
\ee
Thus $6$ is self-conjugate, while $(2,3)$ and $(4,5)$ form conjugate pairs.
Following our general prescription, we first compute the fusion orders of the non-identity basis elements:
\be
d_2=d_3=3,\qquad d_4=d_5=4, \qquad d_6=2. 
\ee

We then assign the non-identity elements to the residues in the ambient cyclic factors. Consider the $\ZZ_4$ cyclic factor, with $L_1=4$, accommodating the circle $\{4,6,5\}$. We assign
\be
r_1(4)=1, \quad r_1(6)=2, \quad r_1(5)=3,
\label{eq:assign_fusion-C9}
\ee
so that the occupied subset is $\Omega_4=\{1,2,3\}$. Since elements $2$ and $3$ do not belong to this circle, we have $r_1(2)=r_1(3)=0$. 
The assignment in Eq.~\eqref{eq:assign_fusion-C9} manifestly satisfies the order-compatibility condition and correctly realizes the conjugation condition, since $4$ and $5$ occupy opposite residues while the self-conjugate element $6$ sits at the order-two element. It also satisfies the pairwise fusion-compatibility conditions:
\begin{align}
1+1 &\equiv 2, &
1+2 &\equiv 3, &
1+3 &\equiv 0 \pmod 4, \\
2+2 &\equiv 0, &
2+3 &\equiv 1, &
3+3 &\equiv 2 \pmod 4,
\end{align}
in agreement with
\begin{align}
6 &\prec 4\otimes4, &
5 &\prec 4\otimes6, &
1 &\prec 4\otimes5, \\
1 &\prec 6\otimes6, &
4 &\prec 6\otimes5, &
6 &\prec 5\otimes5.
\end{align}
Similarly, all remaining modular relations supported on $\Omega_4=\{1,2,3\}$ are compatible with the fusion rules, as verified explicitly in the Supplementary Material. Hence $\{4,6,5\}$ defines a consistent $\mathbb Z_4$ circle. Likewise, \(\{2,3\}\) defines a consistent \(\mathbb Z_3\) circle, with
\begin{equation}
r_2(2)=1,\qquad r_2(3)=2,
\end{equation}
so that $\Omega_3=\{1,2\}$, while $r_2(4)=r_2(5)=r_2(6)=0$.

The two consistent circles above define the lifted Abelian group
\begin{equation}
G_{\mathrm{lift}}=\mathbb Z_4\times \mathbb Z_3.
\end{equation}
The lifted charge of each basis element is its residue vector, namely
\begin{equation}
\begin{aligned}
q(1)&=(0,0),\\
q(2)&=(0,1),\qquad q(3)=(0,2),\\
q(4)&=(1,0),\qquad q(5)=(3,0),\qquad q(6)=(2,0).
\end{aligned}
\end{equation}

Following the general criterion in Eq.~\eqref{eq:allowed-vertex} and the fusion rules in Table~\ref{tab:fusion-C9}, one can construct all vertices allowed by the NISRs. 
This yields roughly fifty renormalizable vertices built from the non-identity elements. Rather than listing all these operators, we focus on the quartic terms 
\be
2366, \quad 2556, \quad 4456, \quad 3456,
\label{eq:input_quartic}
\ee
which illustrate four distinct cases. All four vertices are allowed by the NISRs and must therefore be treated on the same footing from the viewpoint of NISRs. However, their distinction appears only in the lifted description, as seen from the associated spurion charges:
\begin{equation}
\begin{aligned}
q(\lambda_{2366})=(0,0), \quad  q(\lambda_{2556})=(0,2),  \\
q(\lambda_{4456})=(1,0), \quad  q(\lambda_{3456})=(2,1).
\end{aligned}
\end{equation}
The first vertex is neutral under $\ZZ_4\times \ZZ_3$. The second couples elements from different circles and preserves the $\ZZ_4$ factor while breaking the $\ZZ_3$ factor; this cannot arise from interactions among $2$ and $3$ since they are invertible. The third involves only elements in the $\ZZ_4$ factor, yet still breaks $\ZZ_4$; this is possible because fusion products among $4,5,6$ contain multiple channels, which is a hallmark of non-invertible fusion algebras. The fourth combines both features: it couples elements from different circles and breaks both $\ZZ_4$ and $\ZZ_3$.
This example makes explicit that NISR-allowed interactions need not be neutral under $G_{\mathrm{lift}}$, and that the lifted description does not replace the NISRs by exact $G_{\mathrm{lift}}$ invariance; rather, it trades non-invertibility for a structured pattern of $G_{\mathrm{lift}}$-breaking terms.~\footnote{Conversely, not every $G_{\mathrm{lift}}$-breaking term is NISR-allowed. For example, the quartic interactions $2222$ and $2224$ break $G_{\mathrm{lift}}$ but do not satisfy Eq.~\eqref{eq:allowed-vertex}.} 

\begin{figure}[t]
\centering
\includegraphics[scale=0.2,angle=-90]{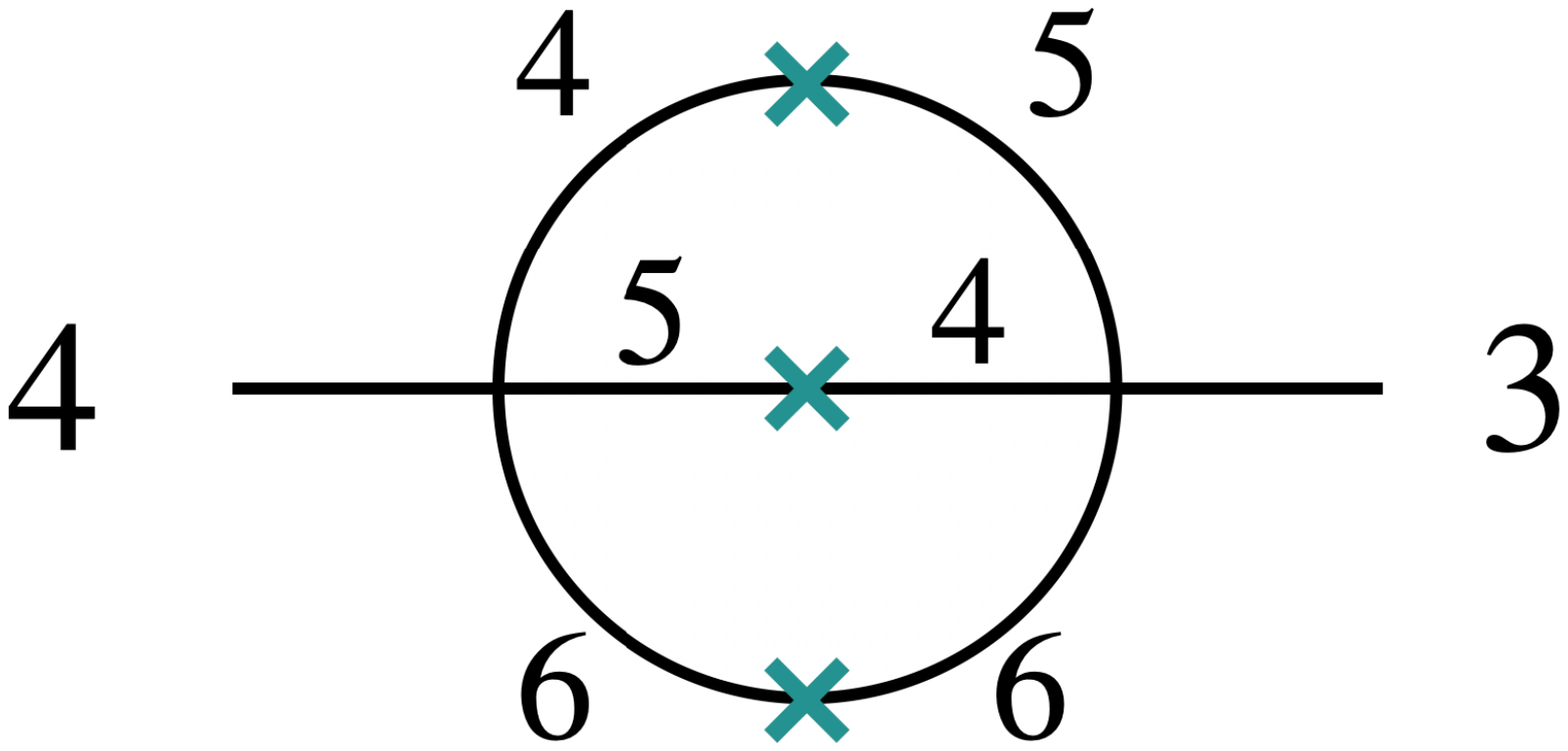}
\caption{Radiative generation of the $34$ term at the two-loop order induced from the two tree-level quartic terms $4456$ and $3456$. The particles are labeled by the elements in the fusion algebra shown in Table~\ref{tab:fusion-C9}. } 
\label{fig:Exp_2loop}
\end{figure}

Let us next consider the two-loop generation of the effective vertex $34$ from the tree-level quartic terms $4456$ and $3456$. This example illustrate how spurion labels are tracked when general amplitudes are constructed from basic building blocks. This is done by contracting the conjugate pairs as shown in Figure~\ref{fig:Exp_2loop}. The induced vertex has the coupling whose charge under the lifted group is
\be
q(\lambda^{(2)}_{34})=(1,0)+(2,1)=(3,1).
\ee
Equivalently, at the level of non-invertible fusion algebra, this is captured by the relation
\be
(5\otimes 2) \prec (5\otimes 6) \otimes (2\otimes 6), 
\ee
where the couplings are labeled by the ``composite spurions''~\cite{Suzuki:2025kxz}:
\begin{equation}
\ell(\lambda^{(2)}_{34})=5\otimes 2,\quad
\ell(\lambda_{4456})=5\otimes 6,\quad
\ell(\lambda_{3456})=2\otimes 6.
\end{equation}

%%%%%%%%%%%%%%%%%%%%%%%%%%%%%%%%%%%%%%%%%%%%%%%%%%%%%%%%%%%%%%%%%%%%%%%%%
\section{Conclusion and outlook}
%%%%%%%%%%%%%%%%%%%%%%%%%%%%%%%%%%%%%%%%%%%%%%%%%%%%%%%%%%%%%%%%%%%%

We have formulated a general spurion prescription for NISRs directly from fusion-algebra data, thereby elevating previously studied special cases to a systematic framework.
In the lifted description, non-invertibility is traded for a structured pattern of \(G_{\mathrm{lift}}\)-breaking terms. 

More broadly, the present work supports the view that NISRs capture a very minimal layer of selection rules governing dynamical particle scatterings. Although such rules may appear exotic from a traditional symmetry perspective, they can be understood as a natural generalization of tensor-product structure with the representation indices stripped away. Precisely because this structure is so minimal, it is generically not rigid under radiative corrections. The spurion framework developed here provides a systematic way to track and organize such breaking.

Looking ahead, it would be interesting to understand the full set of obstructions to loop-induced violation of NISRs, possible ultraviolet completions of such selection rules beyond string theory, and more general classes of selection rules governing particle scatterings, including in the presence of defects.

%%%%%%%%%%%%%%%%%%%%%%%%%%%%%%%%%%%%%%%%%%%%%%%%%%%%%%%%%%%%%%%%%%%%%%%%%
\section{Acknowledgment}
%%%%%%%%%%%%%%%%%%%%%%%%%%%%%%%%%%%%%%%%%%%%%%%%%%%%%%%%%%%%%%%%%%%%%%%%%
I would first like to thank Motoo Suzuki and Hao Y. Zhang for their collaboration on related projects and for their useful discussions. 
I thank Jian Zhou for a comment at the early stage of this project.
I also thank ChatGPT 5.4 Thinking (OpenAI) and Claude Opus 4.6 (Anthropic) for helpful contributions to the computation of the examples and the writing of the paper. L.X.X. is fully responsible for the scientific content of the paper, who guided the AI research assistants and independently verified the results. 
The work of L.X.X. is partially supported by European Research Council (ERC) grant n.101039756. L.X.X. is also grateful to IHES for the hospitality during the workshop ``\href{https://indico.math.cnrs.fr/event/16013/overview}{Symmetry and Topology in Particle Physics}'' while this work was under preparation, and to the participants for useful discussions related to this project. 
%%%%%%%%%%%%%%%%%%%%%%%%%%%%%%%%%%%%%%%%%%%%%%%%%%%%%%%%%%%%%%%%%%%%%

\bibliography{Noninvert_SP.bib}

%%%%%%%%%%%%%%%%%%%%%%%%%%%%%%%%%%%%%%%%

%%%%%%%%%%%%%%%%%%%%%%%%%%%%%%%%%%%%%%%
% Supplementary Material %
%%%%%%%%%%%%%%%%%%%%%%%%%%%%%%%%%%%%%%%%%%%%
\clearpage
\onecolumngrid
\appendix
\makeatletter

\label{supp}

%%%%%%%%%%%%%%%%%%%%%%%%%%%%%%%%%%%%%%%%%%%%%%%%%%%%
\newpage

\begin{center}
   \textbf{\large SUPPLEMENTARY MATERIAL \\[.2cm] ``A General Prescription for Spurion Analysis of Non-Invertible Selection Rules'' }\\[.2cm]
\end{center}
%%%%%%%%%%%%%%%%%%%%%%%%%%%%%%%%%%%%%%%%%%%%%%%%%%%%%

In this supplementary material, we state the full modular consistency condition used in the Step~\emph{(ii) Cyclic reconstruction} of the main text, and summarize in Tables~\ref{tab:KTZ} and~\ref{tab:Dong} the examples analyzed using our general prescription.

%%%%%%%%%%%%%%%%%%%%%%%%%%%%%%%%%%%%%%%%%%%%%%%%%%%
\noindent
\section{Full modular consistency condition}
\label{app:mod}
\setcounter{equation}{0}
\setcounter{figure}{0}
\setcounter{table}{0}
\renewcommand{\theequation}{A\arabic{equation}}
\renewcommand{\thefigure}{A\arabic{figure}}
\renewcommand{\thetable}{A\arabic{table}}
%%%%%%%%%%%%%%%%%%%%%%%%%%%%%%%%%%%%%%%%%%%%%%%%%%%%

Fix one ambient cyclic factor $\mathbb Z_L$, and let
\begin{equation}
\Omega_L\subset \mathbb Z_L\setminus\{0\}
\label{eq:OmegaL-supp}
\end{equation}
denote the occupied subset of nonzero residues. 
We restrict to injective assignments on the occupied subset $\Omega_L$, so that each occupied residue $r\in\Omega_L$ carries a unique basis element, which we denote by $x_r$.

A candidate assignment is consistent only if, for every finite collection of occupied residues
\begin{equation}
r_1,\dots,r_k\in \Omega_L
\label{eq:occupied-residues-supp}
\end{equation}
and every choice of nonnegative integers
\begin{equation}
n_1,\dots,n_k\in \mathbb Z_{\ge 0},
\label{eq:nonnegative-integers-supp}
\end{equation}
the following conditions hold:
\begin{itemize}
\item If
\begin{equation}
n_1 r_1+\cdots+n_k r_k \equiv t \pmod L
\label{eq:mod-rel-nonzero-supp}
\end{equation}
with $t\in \Omega_L$, then
\begin{equation}
x_t \prec x_{r_1}^{\,n_1}\cdots x_{r_k}^{\,n_k}.
\label{eq:full-consistency-nonzero-supp}
\end{equation}

\item If
\begin{equation}
n_1 r_1+\cdots+n_k r_k \equiv 0 \pmod L,
\label{eq:mod-rel-zero-supp}
\end{equation}
then
\begin{equation}
1 \prec x_{r_1}^{\,n_1}\cdots x_{r_k}^{\,n_k}.
\label{eq:full-consistency-zero-supp}
\end{equation}
\end{itemize}
Equations~\eqref{eq:full-consistency-nonzero-supp} and \eqref{eq:full-consistency-zero-supp} define the full modular consistency condition. They make explicit that cyclic reconstruction is not merely a pairwise matching problem: one must require compatibility between fusion and \emph{all} modular relations supported on $\Omega_L$.

At first sight, the above condition appears infinite because it involves arbitrary nonnegative integers $n_i$. In practice, however, the consistency problem is finite. On the arithmetic side, only the residue class
\begin{equation}
n_1 r_1+\cdots+n_k r_k \pmod L
\label{eq:residue-class-supp}
\end{equation}
matters, and there are only finitely many residue classes modulo $L$. Moreover, since $N^z_{x,y}\geq 0$ and $1\prec x^{d_{x_{r_i}}}_{r_i}$, it follows that 
\be
x^{n_i}_{r_i}\prec x^{n_i+m \ d_{x_{r_i}}}_{r_i}, \quad m\in \mathbb{Z}_{\geq 0}.
\ee
Thus, it is sufficient to consider exponents modulo $d_{x_{r_i}}$ for each $x_{r_i}$. (Since $d_{x_{r_i}}$ divides $L$, one can also check exponents modulo $L$ for all basis elements in the same $\ZZ_L$ cyclic factor.)
Since $\Omega_L$ is finite, only finitely many modular relations among occupied residues can arise.

In practice, one may proceed more concretely as follows:
\begin{enumerate}
\item Choose a candidate ambient order $L$. If several basis elements are to be accommodated in the same cyclic factor, then $L$ must be a common multiple of their fusion orders; a natural starting point is therefore the least common multiple of the relevant $d_x$.

\item Choose a candidate occupied subset $\Omega_L\subset \mathbb Z_L\setminus\{0\}$ and begin a residue assignment consistent with the fusion-order data. If the factor contains an element of fusion order $L$, then, up to cyclic relabeling, one may place one such element at residue $1$, thereby fixing the overall origin of the cyclic labeling. More generally, one may choose any convenient representative assignment as a starting point.

\item Use the pairwise fusion relations as preliminary constraints to determine or restrict the positions of other candidate elements. In particular, pairwise relations provide an efficient first pass for excluding incompatible assignments.

\item Finally, verify that \emph{all} modular relations among occupied residues are compatible with the fusion rules through Eqs.~\eqref{eq:mod-rel-nonzero-supp}--\eqref{eq:full-consistency-zero-supp}.
\end{enumerate}

In simple examples, the occupied residues may already be fixed by the fusion-order data together with pairwise and low-power checks. Conceptually, however, the correct acceptance criterion is the full modular consistency condition stated above. In more complicated cases, this finite consistency problem is naturally handled by a computer search over the remaining candidate assignments.

To illustrate the full modular consistency condition, we consider the \(\mathbb Z_4\) circle \(\{4,6,5\}\) in the fusion algebra of Table~\ref{tab:fusion-C9}, with residue assignment
\begin{equation}
r(4)=1,\qquad r(6)=2,\qquad r(5)=3
\qquad (\mathrm{mod}\ 4).
\end{equation}
We fix \(n_1=1\) and scan \(n_2,n_3\in\{0,1,2,3\}\), corresponding to monomials
\begin{equation}
4\,6^{n_2}5^{n_3}.
\end{equation}
The relevant modular relation is
\begin{equation}
t \equiv 1+2n_2+3n_3 \pmod 4.
\end{equation}
For \(t=1,2,3\), the consistency condition requires \(x_t\prec 4\,6^{n_2}5^{n_3}\), where
\begin{equation}
x_1=4,\qquad x_2=6,\qquad x_3=5,
\end{equation}
while for \(t=0\) it requires
\begin{equation}
1\prec 4\,6^{n_2}5^{n_3}.
\end{equation}
The representative slice of \(n_1=1\) is summarized in Table~\ref{tab:C9-modular-slice}. As expected, all entries satisfy the required modular consistency condition. The remaining slices of the full modular scan can be worked out analogously.

\begin{table}[htbp]
\centering
\caption{Representative slice of the full modular consistency check for the \(\mathbb Z_4\) circle \(\{4,6,5\}\), with \(n_1=1\) fixed.}
\label{tab:C9-modular-slice}
\renewcommand{\arraystretch}{1.15}
\setlength{\tabcolsep}{5pt}
\footnotesize
\begin{tabular}{cccccl}
\hline
\(n_2\) & \(n_3\) & \(t\) & Required elements & Fusion monomial & Check \\
\hline
0 & 0 & 1 & \(4\) & \(4\) & trivial \\
0 & 1 & 0 & \(1\) & \(4\otimes 5\) & \(1\prec 4\otimes 5\) \\
0 & 2 & 3 & \(5\) & \(4\otimes 5^2\) & \(6\prec 5^2,\; 5\prec 4\otimes 6\) \\
0 & 3 & 2 & \(6\) & \(4\otimes 5^3\) & \(6\prec 5^2,\; 4\prec 5\otimes 6,\; 6\prec 4^2\) \\
\hline
1 & 0 & 3 & \(5\) & \(4\otimes 6\) & \(5\prec 4\otimes 6\) \\
1 & 1 & 2 & \(6\) & \(4\otimes 6\otimes 5\) & \(1\prec 4\otimes 5,\; 6=1\otimes 6\) \\
1 & 2 & 1 & \(4\) & \(4\otimes 6\otimes 5^2\) & \(6\prec 5^2,\; 1\prec 6^2,\; 4=1\otimes 4\) \\
1 & 3 & 0 & \(1\) & \(4\otimes 6\otimes 5^3\) & \(6\prec 5^2,\; 1\prec 6^2,\; 1\prec 4\otimes 5\) \\
\hline
2 & 0 & 1 & \(4\) & \(4\otimes 6^2\) & \(1\prec 6^2,\; 4=1\otimes 4\) \\
2 & 1 & 0 & \(1\) & \(4\otimes 6^2\otimes 5\) & \(1\prec 6^2,\; 1\prec 4\otimes 5\) \\
2 & 2 & 3 & \(5\) & \(4\otimes 6^2\otimes 5^2\) & \(1\prec 6^2,\; 6\prec 5^2,\; 5\prec 4\otimes 6\) \\
2 & 3 & 2 & \(6\) & \(4\otimes 6^2\otimes 5^3\) & \(1\prec 6^2,\; 6\prec 5^2,\; 4\prec 5\otimes 6,\; 6\prec 4^2\) \\
\hline
3 & 0 & 3 & \(5\) & \(4\otimes 6^3\) & \(1\prec 6^2,\; 5\prec 4\otimes 6\) \\
3 & 1 & 2 & \(6\) & \(4\otimes 6^3\otimes 5\) & \(1\prec 6^2,\; 1\prec 4\otimes 5,\; 6=1\otimes 6\) \\
3 & 2 & 1 & \(4\) & \(4\otimes 6^3\otimes 5^2\) & \(1\prec 6^2,\; 6\prec 5^2,\; 1\prec 6^2,\; 4=1\otimes 4\) \\
3 & 3 & 0 & \(1\) & \(4\otimes 6^3\otimes 5^3\) & \(1\prec 6^2,\; 6\prec 5^2,\; 1\prec 6^2,\; 1\prec 4\otimes 5\) \\
\hline
\end{tabular}
\end{table}

\newpage

%%%%%%%%%%%%%%%%%%%%%%%%%%%%%%%%%%%%%%%%%%%%%%%%%%%%%%%%%%%%%%%%%%%%%%%%%%%%%%%
\noindent
\section{Fusion algebras analyzed with the general prescription}
\label{app:exp}
\setcounter{equation}{0}
\setcounter{figure}{0}
\setcounter{table}{0}
\renewcommand{\theequation}{B\arabic{equation}}
\renewcommand{\thefigure}{B\arabic{figure}}
\renewcommand{\thetable}{B\arabic{table}}
%%%%%%%%%%%%%%%%%%%%%%%%%%%%%%%%%%%%%%%%%%%%%%%%%%%%%%%%%%%%%%%%%%%%%%%%%%%%%%%

%
\begin{table*}[htbp]
\centering
\caption{Summary of the 18 fusion algebras in KTZ paper's Appendix~C~\cite{Kaidi:2024wio} (with C.4 excluded) analyzed using the general prescription. 
The columns list the algebra, rank, conjugation action on the non-identity basis elements, fusion orders $d_x$, cyclic reconstruction, and the resulting lifted Abelian group $G_{\mathrm{lift}}$. Here $\bar X = Y $ denotes a conjugate pair and ``s.c.'' denotes a self-conjugate element. Algebras for which all non-identity basis elements are self-conjugate are highlighted in blue. }
\label{tab:KTZ}
\renewcommand{\arraystretch}{1.2}
\footnotesize
\resizebox{\textwidth}{!}{%
\begin{tabular}{cccccc}
\hline
\textbf{Algebra} & \textbf{Rank} & \textbf{Conj.} & \textbf{Orders} & \textbf{Circles} & $\boldsymbol{G}_{\text{lift}}$ \\
\hline
 
  KTZ C.1a & 4 & $\bar 2{=}3; 4\text{ s.c.}$ & $(3,3,2)$ & $\{2,3\}, \{4\}$ & $\ZZ_{3}\times \ZZ_{2}$ \\
\rowcolor{blue!12}
 $\text{PSU}(2)_6\cong \text{HI}(\ZZ_2)$ & 4 & $2\text{ s.c.}; 3\text{ s.c.}; 4\text{ s.c.}$ & $(2,2,2)$ & $\{2\}, \{3\}, \{4\}$ & $\ZZ_{2}\times \ZZ_{2}\times \ZZ_{2}$ \\
 
\rowcolor{blue!12}
 KTZ C.2 & 5 & $2\text{ s.c.}; 3\text{ s.c.}; 4\text{ s.c.}; 5\text{ s.c.}$ & $(2,2,2,2)$ & $\{2\}, \{3\}, \{4\}, \{5\}$ & $\ZZ_{2}\times \ZZ_{2}\times \ZZ_{2}\times \ZZ_{2}$ \\
\rowcolor{blue!12}
 $\text{PSU}(2)_{10}$ & 6 & $2\text{ s.c.}; 3\text{ s.c.}; 4\text{ s.c.}; 5\text{ s.c.}; 6\text{ s.c.}$ & $(2,2,2,2,2)$ & $\{2\}, \{3\}, \{4\}, \{5\}, \{6\}$ & $\ZZ_{2}\times \ZZ_{2}\times \ZZ_{2}\times \ZZ_{2}\times \ZZ_{2}$ \\
 
 KTZ C.5 & 6 & $\bar 2{=}3; \bar 4{=}5; 6\text{ s.c.}$ & $(4,4,3,3,2)$ & $\{2,6,3\}, \{4,5\}$ & $\ZZ_{4}\times \ZZ_{3}$ \\
\rowcolor{gray!6}
 KTZ C.6 & 6 & $\bar 2{=}3; 4\text{ s.c.}; 5\text{ s.c.}; 6\text{ s.c.}$ & $(4,4,2,2,2)$ & $\{2,6,3\}, \{4\}, \{5\}$ & $\ZZ_{4}\times \ZZ_{2}\times \ZZ_{2}$ \\
 
 KTZ C.7 & 6 & $\bar 2{=}3; 4\text{ s.c.}; 5\text{ s.c.}; 6\text{ s.c.}$ & $(3,3,2,2,2)$ & $\{2,3\}, \{4\}, \{5\}, \{6\}$ & $\ZZ_{3}\times \ZZ_{2}\times \ZZ_{2}\times \ZZ_{2}$ \\
\rowcolor{blue!12}
 KTZ C.8 & 6 & $2\text{ s.c.}; 3\text{ s.c.}; 4\text{ s.c.}; 5\text{ s.c.}; 6\text{ s.c.}$ & $(2,2,2,2,2)$ & $\{2\}, \{3\}, \{4\}, \{5\}, \{6\}$ & $\ZZ_{2}\times \ZZ_{2}\times \ZZ_{2}\times \ZZ_{2}\times \ZZ_{2}$ \\
 
 KTZ C.9 & 6 & $\bar 2{=}3; \bar 4{=}5; 6\text{ s.c.}$ & $(3,3,4,4,2)$ & $\{4,6,5\}, \{2,3\}$ & $\ZZ_{4}\times \ZZ_{3}$ \\
\rowcolor{blue!12}
 KTZ C.10 & 6 & $2\text{ s.c.}; 3\text{ s.c.}; 4\text{ s.c.}; 5\text{ s.c.}; 6\text{ s.c.}$ & $(2,2,2,2,2)$ & $\{2\}, \{3\}, \{4\}, \{5\}, \{6\}$ & $\ZZ_{2}\times \ZZ_{2}\times \ZZ_{2}\times \ZZ_{2}\times \ZZ_{2}$ \\
 
 KTZ C.11 & 6 & $\bar 2{=}3; \bar 4{=}5; 6\text{ s.c.}$ & $(3,3,3,3,2)$ & $\{2,3\}, \{4,5\}, \{6\}$ & $\ZZ_{3}\times \ZZ_{3}\times \ZZ_{2}$ \\
\rowcolor{gray!6}
 KTZ C.12 & 6 & $\bar 2{=}3; 4\text{ s.c.}; 5\text{ s.c.}; 6\text{ s.c.}$ & $(3,3,2,2,2)$ & $\{2,3\}, \{4\}, \{5\}, \{6\}$ & $\ZZ_{3}\times \ZZ_{2}\times \ZZ_{2}\times \ZZ_{2}$ \\
 
\rowcolor{blue!12}
 KTZ C.13 & 6 & $2\text{ s.c.}; 3\text{ s.c.}; 4\text{ s.c.}; 5\text{ s.c.}; 6\text{ s.c.}$ & $(2,2,2,2,2)$ & $\{2\}, \{3\}, \{4\}, \{5\}, \{6\}$ & $\ZZ_{2}\times \ZZ_{2}\times \ZZ_{2}\times \ZZ_{2}\times \ZZ_{2}$ \\
\rowcolor{gray!6}
 KTZ C.14 & 7 & $\bar 2{=}3; 4\text{ s.c.}; 5\text{ s.c.}; 6\text{ s.c.}; 7\text{ s.c.}$ & $(4,4,2,2,2,2)$ & $\{2,7,3\}, \{4\}, \{5\}, \{6\}$ & $\ZZ_{4}\times \ZZ_{2}\times \ZZ_{2}\times \ZZ_{2}$ \\
 
\rowcolor{blue!12}
 KTZ C.15 & 7 & $2\text{ s.c.}; 3\text{ s.c.}; 4\text{ s.c.}; 5\text{ s.c.}; 6\text{ s.c.}; 7\text{ s.c.}$ & $(2,2,2,2,2,2)$ & $\{2\}, \{3\}, \{4\}, \{5\}, \{6\}, \{7\}$ & $\ZZ_{2}\times \ZZ_{2}\times \ZZ_{2}\times \ZZ_{2}\times \ZZ_{2}\times \ZZ_{2}$ \\
\rowcolor{blue!12}
 KTZ C.16 & 7 & $2\text{ s.c.}; 3\text{ s.c.}; 4\text{ s.c.}; 5\text{ s.c.}; 6\text{ s.c.}; 7\text{ s.c.}$ & $(2,2,2,2,2,2)$ & $\{2\}, \{3\}, \{4\}, \{5\}, \{6\}, \{7\}$ & $\ZZ_{2}\times \ZZ_{2}\times \ZZ_{2}\times \ZZ_{2}\times \ZZ_{2}\times \ZZ_{2}$ \\
 
\rowcolor{blue!12}
 KTZ C.17 & 7 & $2\text{ s.c.}; 3\text{ s.c.}; 4\text{ s.c.}; 5\text{ s.c.}; 6\text{ s.c.}; 7\text{ s.c.}$ & $(2,2,2,2,2,2)$ & $\{2\}, \{3\}, \{4\}, \{5\}, \{6\}, \{7\}$ & $\ZZ_{2}\times \ZZ_{2}\times \ZZ_{2}\times \ZZ_{2}\times \ZZ_{2}\times \ZZ_{2}$ \\
\rowcolor{blue!12}
 KTZ C.18 & 7 & $2\text{ s.c.}; 3\text{ s.c.}; 4\text{ s.c.}; 5\text{ s.c.}; 6\text{ s.c.}; 7\text{ s.c.}$ & $(2,2,2,2,2,2)$ & $\{2\}, \{3\}, \{4\}, \{5\}, \{6\}, \{7\}$ & $\ZZ_{2}\times \ZZ_{2}\times \ZZ_{2}\times \ZZ_{2}\times \ZZ_{2}\times \ZZ_{2}$ \\
 
\hline
\end{tabular}
}
\end{table*}

\begin{table*}[htbp]
\centering
\caption{Summary of selected fusion algebras of DJKNO~\cite{Dong:2025jra}, including examples from discrete gauging and group conjugacy classes, analyzed using the general prescription. 
The columns list the algebra, rank, conjugation action on the non-identity basis elements, fusion orders $d_x$, cyclic reconstruction, and the resulting lifted Abelian group $G_{\mathrm{lift}}$. Here $\bar X = Y $ denotes a conjugate pair and ``s.c.'' denotes a self-conjugate element. Algebras for which all non-identity basis elements are self-conjugate are highlighted in blue.}
\label{tab:Dong}
\renewcommand{\arraystretch}{2.5}
\setlength{\tabcolsep}{2pt}
\footnotesize
\begin{tabular}{cccccc}
\hline
\textbf{Algebra} & \textbf{Rank} & \textbf{Conj.} & \textbf{Orders} & \textbf{Circles} & $\boldsymbol{G}_{\textbf{lift}}$ \\
\hline
$\ZZ_3$ gauging of $\ZZ_7$ & 3 & $\bar C_3^1{=}C_3^2$ & $(3,3)$ & $\{C_3^1,C_3^2\}$ & $\ZZ_{3}$ \\
\rowcolor{gray!6}
Conjugacy classes of $T_7$ & 5 & \shortstack[c]{$\bar C_3^1{=}C_3^2$\\[3pt]$\bar C_7^{(1)}{=}C_7^{(2)}$} & $(3,3,3,3)$ & \shortstack[c]{$\{C_3^1,C_3^2\},$\\[3pt]$\{C_7^{(1)},C_7^{(2)}\}$} & $\ZZ_{3}\times \ZZ_{3}$ \\
\rowcolor{blue!12}
$\ZZ_3$ gauging of $\ZZ_2{\times}\ZZ_2'$ & 2 & $C_3^{(1,0)}$ s.c. & $(2)$ & $\{C_3^{(1,0)}\}$ & $\ZZ_{2}$ \\
$\ZZ_3$ gauging of $\ZZ_3{\times}\ZZ_3'$ & 5 & \makecell{$\bar C_1^1{=}C_1^2$\\[3pt]$\bar C_3^{(1,0)}{=}C_3^{(2,0)}$} & $(3,3,3,3)$ & \makecell{$\{C_1^1,C_1^2\},$\\[3pt]$\{C_3^{(1,0)},C_3^{(2,0)}\}$} & $\ZZ_{3}\times \ZZ_{3}$ \\
\rowcolor{gray!6}
$\ZZ_3$ gauging of $\ZZ_4{\times}\ZZ_4'$ & 6 & \shortstack[c]{$\bar C_3^{(1,0)}{=}C_3^{(3,0)}$\\[3pt]$C_3^{(2,0)}$ s.c.\\[3pt]$\bar C_3^{(1,2)}{=}C_3^{(2,1)}$} & $(3,2,3,3,3)$ & \shortstack[c]{$\{C_3^{(1,0)},C_3^{(3,0)}\},$\\[3pt]$\{C_3^{(2,0)}\},$\\[3pt]$\{C_3^{(1,2)},C_3^{(2,1)}\}$} & $\ZZ_{3}\times \ZZ_{2}\times \ZZ_{3}$ \\
\rowcolor{blue!12}
$\ZZ_3$ gauging of $\ZZ_2 \times \ZZ_2 \times \ZZ_2$ & 4 & \shortstack[c]{$C_1^1,\;C_3^{(1,0,0)},$\\[3pt]$C_3^{(1,1,0)}$ s.c.} & $(2,2,2)$ & \shortstack[c]{$\{C_1^1\},\;\{C_3^{(1,0,0)}\},$\\[3pt]$\{C_3^{(1,1,0)}\}$} & $\ZZ_2 \times \ZZ_2 \times \ZZ_2$ \\
Conjugacy classes of $\Sigma(24)$ & 8 & \makecell{$C_1^1,\;C_3^{(1,1,0)},\;C_3^{(1,0,0)}$ s.c.\\[3pt]$\bar B_4^{(1,0,0,0)}{=}B_4^{(2,0,0,0)}$\\[3pt]$\bar B_4^{(1,1,0,0)}{=}B_4^{(2,1,0,0)}$} & \makecell{$(2,2,2,$\\[3pt]$3,6,3,6)$} & \makecell{$\{C_1^1,B_4^{(1,0,0,0)},B_4^{(2,0,0,0)}\},$\\[3pt]$\{C_3^{(1,0,0)},B_4^{(1,1,0,0)},B_4^{(2,1,0,0)}\},$\\[3pt]$\{C_3^{(1,1,0)}\}$} & $\ZZ_{6}\times \ZZ_{6}\times \ZZ_{2}$ \\
\rowcolor{blue!12}
$S_3$ gauging of $\ZZ_2{\times}\ZZ_2'$ & 2 & $C_3^1$ s.c. & $(2)$ & $\{C_3^1\}$ & $\ZZ_{2}$ \\
$S_3$ gauging of $\ZZ_3{\times}\ZZ_3'$ & 4 & \makecell{$\bar C_1^1{=}C_1^2$\\[3pt]$C_6^{(1,0)}$ s.c.} & $(3,3,2)$ & $\{C_1^1,C_1^2\},\;\{C_6^{(1,0)}\}$ & $\ZZ_{3}\times \ZZ_{2}$ \\
\rowcolor{gray!6}
$\ZZ_3$ gauging of $\ZZ_9$ & 5 & \shortstack[c]{$\bar C_1^3{=}C_1^6$\\[3pt]$\bar C_3^{(1)}{=}C_3^{(2)}$} & $(3,3,3,3)$ & \shortstack[c]{$\{C_1^3,C_1^6\},$\\[3pt]$\{C_3^{(1)},C_3^{(2)}\}$} & $\ZZ_{3}\times \ZZ_{3}$ \\
Conjugacy classes of $\Delta(12)\cong A_4$ & 4 & \makecell{$C_3^{(1,0)}$ s.c.\\[3pt]$\bar C_4^1{=}C_4^2$} & $(2,3,3)$ & $\{C_3^{(1,0)}\},\;\{C_4^1,C_4^2\}$ & $\ZZ_{2}\times \ZZ_{3}$ \\
\rowcolor{blue!12}
Conjugacy classes of $\Delta(24)\cong S_4$ & 5 & \shortstack[c]{$C_3^{(1,0)},\;C_8,$\\[3pt]$B_6^{(0)},\;B_6^{(1)}$ s.c.} & $(2,2,2,2)$ & \shortstack[c]{$\{C_3^{(1,0)}\},\;\{C_8\},$\\[3pt]$\{B_6^{(0)}\},\;\{B_6^{(1)}\}$} & $\ZZ_{2}\times \ZZ_{2} \times \ZZ_{2} \times \ZZ_{2}$ \\
\hline
\end{tabular}
\end{table*}

\end{document}